\documentclass[%
 aip,
 amsmath,amssymb,
 reprint,%
]{revtex4-1}

\usepackage{graphicx}
\usepackage{dcolumn}
\usepackage{bm}

\usepackage[utf8]{inputenc}
\usepackage[T1]{fontenc}
\usepackage{mathptmx}
\usepackage{etoolbox}

\makeatletter
\def\@email#1#2{%
 \endgroup
 \patchcmd{\titleblock@produce}
  {\frontmatter@RRAPformat}
  {\frontmatter@RRAPformat{\produce@RRAP{*#1\href{mailto:#2}{#2}}}\frontmatter@RRAPformat}
  {}{}
}%
\makeatother
\begin{document}

\preprint{AIP/123-QED}

\title[Polarization state control]{Polarization state control for high peak power applications}
\author{E.~M.~Starodubtseva}
 \affiliation{Faculty of Physics, M.V. Lomonosov Moscow State University, Moscow, Russia}
\email{starodubtceva.em19@physics.msu.ru}
\author{I.~N.~Tsymbalov}%
\affiliation{Faculty of Physics, M.V. Lomonosov Moscow State University, Moscow, Russia}
\affiliation{Institute for Nuclear Research of the Russian Academy of Sciences, Moscow, Russia}
\author{D.~A.~Gorlova}%
\affiliation{Faculty of Physics, M.V. Lomonosov Moscow State University, Moscow, Russia}
\affiliation{Institute for Nuclear Research of the Russian Academy of Sciences, Moscow, Russia}
\author{K.~A.~Ivanov}%
\affiliation{Faculty of Physics, M.V. Lomonosov Moscow State University, Moscow, Russia}
\affiliation{P.N. Lebedev Physical Institute, Moscow, Russia}
\author{A.~B.~Savel'ev}%
\affiliation{Faculty of Physics, M.V. Lomonosov Moscow State University, Moscow, Russia}
\affiliation{P.N. Lebedev Physical Institute, Moscow, Russia}

\date{\today}

\begin{abstract}

Numerous applications in the extreme field science are possible with circularly polarized high peak power ultrashort pulses. Commonly used quarter wave plates are not applicable here, while multi-mirror schemes are very complicated. We showed that the simple PET film $\approx$20 $\mu$m thick can be used to control the polarization state of the high peak power beam and achieve ellipticity of $\approx$0.8 with negligible nonlinear phase distortion. The film can withstand $10^3$ shots at intensity of $I = 3 \cdot 10^{12}$ W/cm$^2$ without visible damage. Thus, for a PW laser system and beam diameter of 20 cm the PET film can be used for quite a long time. We proved this experimentally  using 1 TW femtosecond Ti:Sa laser  measuring the angular distribution of the second harmonic from the plasma channel created in an undercritical gas plume.

\end{abstract}

\maketitle
\section{Introduction}

Polarization state is crucial for high-intensity laser matter interaction \cite{doi:10.1103/RevModPhys.78.309, doi:10.1140/epja/s10050-023-01043-2}. Most concepts in extreme field physics rely on the linearly polarized light since it is the natural state for an ultrashort laser pulse after its amplification. 
The circularly polarized (CP) laser radiation provides new features and regimes here, but one needs sophisticated setup to produce CP high peak power ultrashort pulses. 

Nevertheless  a lot of numerical and analytical papers suggested CP pulses for extreme laser-matter interaction. 
CP creates favorable conditions for stable acceleration of ions by radiation pressure of ultra-powerful laser pulses (RPA scheme) \cite{doi:10.1016/j.hedp.2012.02.002, doi:10.1103/PhysRevLett.103.245003, doi:10.1063/1.4958654}. The electron heating is lower and thus  plasma opacity preserves in this regime providing much faster growth of energy of accelerated ions with laser intensity ($\propto I^2$) \cite{doi:10.1063/1.4958654}.

The CP is also important  for the laser-plasma electron acceleration \cite{doi:10.1038/nphys966}. Utilization of a CP laser pulse is more efficient than a linearly polarized pulse for electrons acceleration by the ponderomotive force due to the fact that it is greater for circular polarization than for linear polarization by a factor of 2, so the threshold intensity needed for acceleration is lower for the former. Some PIC-simulations results demonstrate improvement of an electron beam quality in laser wakefield acceleration by a CP laser pulse \cite{doi:10.1088/1361-6587/abfd7d}. The CP causes modification of initial conditions for the electron trajectories that leads to significant increase of self-injection efficiency.

In \cite{doi:10.1063/1.1765656, doi:10.1063/1.1430436} it was demonstrated analytically  that during the interaction of plasma electrons in the ion channel with a CP laser pulse they rotate around the propagation direction of the laser pulse and create an azimuthal current that in its turn generates an axial magnetic field. The resonance occurs for the two optimal magnetic fields, resulting in an efficient energy exchange between accelerated electrons and  electric field of the laser pulse. The method above is actually the hot electron version of inverse Faraday effect (IFE) in a plasma using a self-channeling laser pulse of relativistic intensity \cite{doi:10.1103/PhysRevLett.87.215004}. This method, using high-intensity CP laser pulses, produces  even gigagaus \cite{doi:10.1088/1367-2630/ac0573, doi:10.1038/s41598-023-28753-3} magnetic fields.

High peak power CP (or even elliptically polarized ones) laser pulses are also considered for positron beams production from interaction with thin foils or thin undercritical plasma slabs 
\cite{doi:10.1103/PhysRevD.92.085001,
doi:10.1063/5.0104670}.  Comparing to a linearly polarized pulse, the circularly
polarized ones are in favor since the transverse
$JxB$ heating is absent, and electrons/positrons will
rotate around the laser axis. Here the scheme exploiting colliding CP pulses was demonstrated to be an effective way for the Breit–Wheeler positron beam generation. 

However, standard methods of polarization state control are not well applicable due to the Kerr effect of high intensity laser pulses inside a quarter (or half)-wave plate (besides of the group velocity dispersion). For example, a standard zero order $\lambda/4$ quartz plate (6.4 mm thick) for a 1 TW laser at 800 nm with beam diameter of 1 cm produces the nonlinear phase shift $B = \frac{2\pi n}{\lambda}\int n_2 I(z) dz$ (where $\lambda$ -- laser wavelength, $n=1.45, n_2=2.7\cdot10^{-20}~\frac{m^2}{W}$ -- linear and nonlinear refractive indexes, $I(z)$ -- the optical intensity at the beam axis $z$) equals to approximately 20, which indicates strong self-action. Alternatively the CP light at high peak powers can be obtained using multi-mirror configuration \cite{doi:10.1364/AO.16.001082,doi:10.1364/OE.20.020742} or splitting the linearly polarized light by two beams, rotating the polarization of one beam with a mirror by 90$^o$ and recombining the beams. The both approaches are very complicated technically, especially with PW laser beams.

Possible alternative represents by a birefringent thin films.  The nature of the birefringence in some films, such as cellophane, was described by Feynman  \cite{Feynman}. It comes from the molecules alignment  along one direction due to manufacturing procedures inside the film, that causes a difference in the refractive index along and athwart to this direction.  Various materials were used in laser experiments to make circular polarization from the linear one such as  polyester \cite{doi:10.1364/AO.46.005129}, lactic acid (chiral material)  \cite{doi:10.3934/bioeng.2022024}, and plastic \cite{doi:10.1016/S0040-6090(97)01001-8}. 
However, all the materials described have a relatively low damage threshold, i.e. will not capable to withstand a high radiation fluence. 

In this work we present another material -- polyethylene terephthalate (PET) that is able to withstand a TW peak power of laser radiation. We characterize the PET film $21~\mu m$ thick as polarization control unit and demonstrate how the large surface area  PET film allow to make experiments with TW/cm$^2$ laser radiation with controlled polarization state. We exemplified this in the dedicated experiment on second harmonic generation from the walls of plasma channel in subcritical plasma created by the tightly focused relativistic femtosecond laser pulse with linear or circular polarization.

\section{Changes in the polarization state by the PET film}

\begin{figure}
\centering
\includegraphics[scale=0.23]{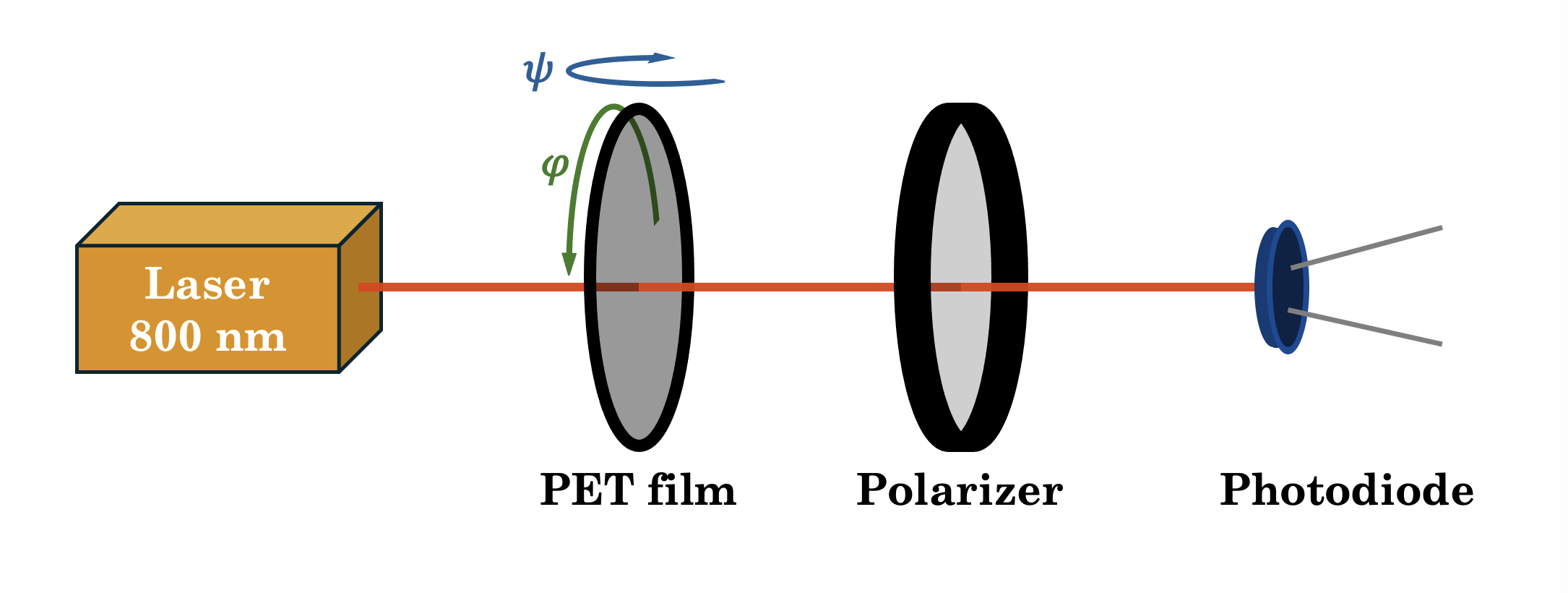}
\caption{\label{fig0} Experimental setup. }
\end{figure}

\begin{figure}[h]
\centering
\includegraphics[scale=0.25]{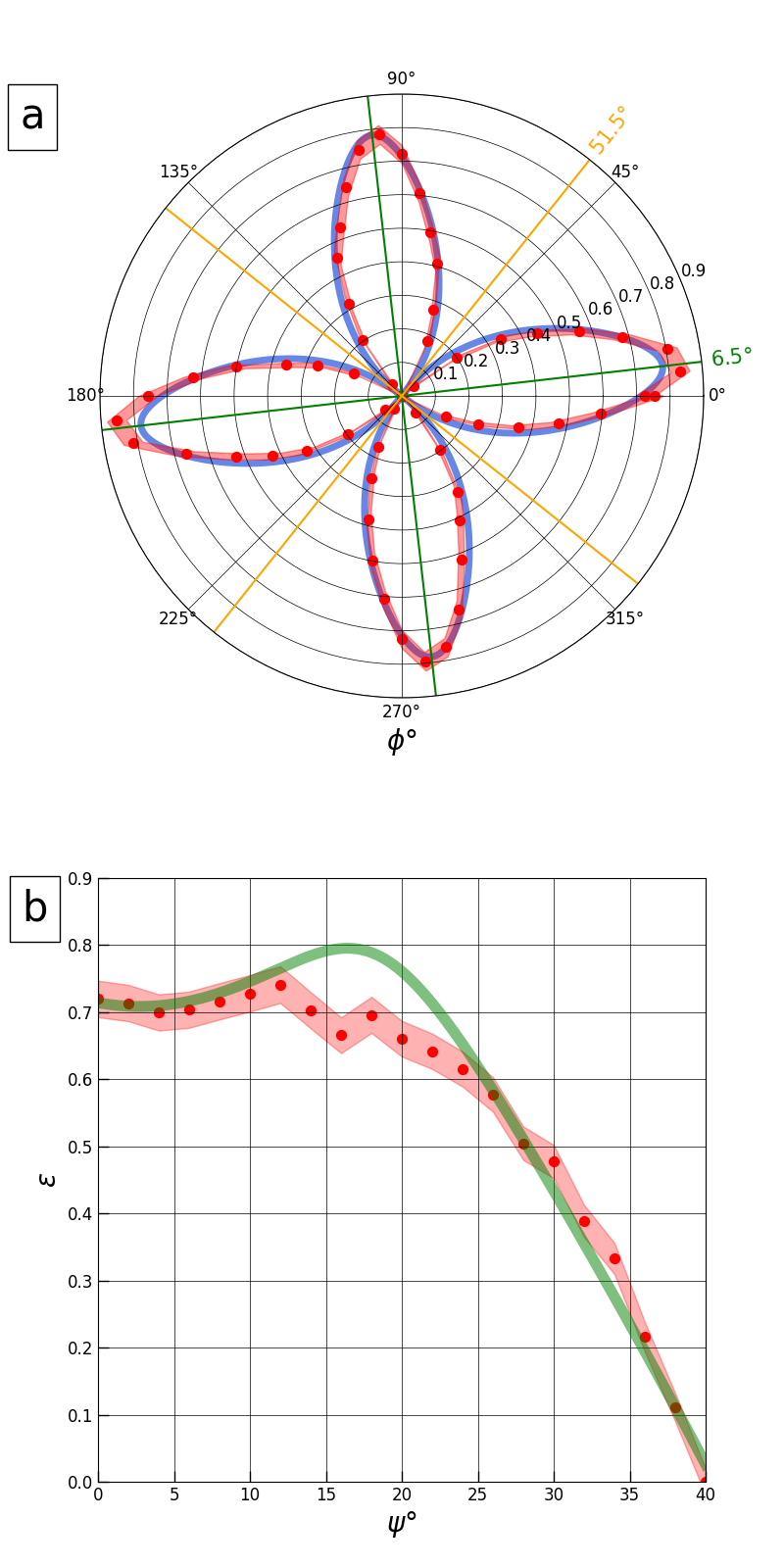}
\caption{\label{fig1} The polarization ellipticity $\epsilon$ dependence on the rotation angle of the film: in the vertical plane (a),  $\phi$ ($\psi=0^{\circ}$ ): red -- experimental results, blue - theoretical dependence (explained in the text), green line depicts maximum ellipticity direction, orange lines -- film axes (zero ellipticity) direction, and in the horizontal plane (b), $\psi$ ($\phi=0^{\circ}$): red -- experimental results, green -- approximation.}
\end{figure}

We used linear polarized (vertical) continuous radiation from the diode laser at 800 nm to study the polarization changes  after the PET film. The radiation passed through a rotating polarizer-analyzer after the film and then it was recorded by the photodiode. (see fig. \ref{fig0}). Fig. \ref{fig1} represents the dependence of the laser radiation ellipticity $\epsilon=\sqrt{\frac{I_{min}}{I_{max}}}$ after the film  on the rotation angles $\phi$  in the vertical plane and $\psi$ in the horizontal plane.

According to the experimental dependence of ellipticity $\epsilon$ on the angle of rotation in the vertical plane (see fig. \ref{fig1}a, red line), the vertical direction of the film axis (zero ellipticity) was obtained at $\phi_{axis}=51.5^{\circ}$ (see orange line in fig. \ref{fig1}a). This value was used for approximation of the $\epsilon(\psi)$ (see. fig \ref{fig1}b), that takes into account rotation of the refractive indices ellipse and increase in the film thickness. This provides with  angle of the horizontal direction of the film axis $\psi_{axis}=1.4349^{\circ}$, parameters of the refractive indices ellipse: $n_o=1.44599$ -- ordinary wave refractive index, $n_e=1.49618$ -- maximum value of extraordinary wave refractive index and the film thickness $d=20.553~\mu m$.
The phase shift $\delta$ was calculated, $\delta = 2\pi\frac{(n_e-n_o)d}{\lambda}=2.58\pi$, using the estimated values. Further, the analytical dependence of the ellipticity $\epsilon$  on the angle $\phi$ was plotted in fig. \ref{fig1}a (blue curve). One can see that the theoretical dependence accurately describes the experimental data.
Note also, that transmission  of the film $I/I_0$ changes from 0.70 to 0.78 with $\psi$ changing from $-40^{\circ}$ to $40^{\circ}$ following Fresnel's equation for the thin film. 

In order to analyze the homogeneity of the polarization properties of the film within the typical size of the laser beam, the polarization ellipticity $\epsilon$ for normal incident at different points in the film plane within centimeter scale was measured. The $\epsilon$ value is stable on the scale of centimeters ($\epsilon=0.76\pm0.03$). This allows using such a film to change polarization state of large aperture beams from high intensity, petawatt laser facilities. 

\section{PET film utilisation with femtosecond laser pulses}

\begin{figure}[h]
\centering
\includegraphics[scale=0.25]{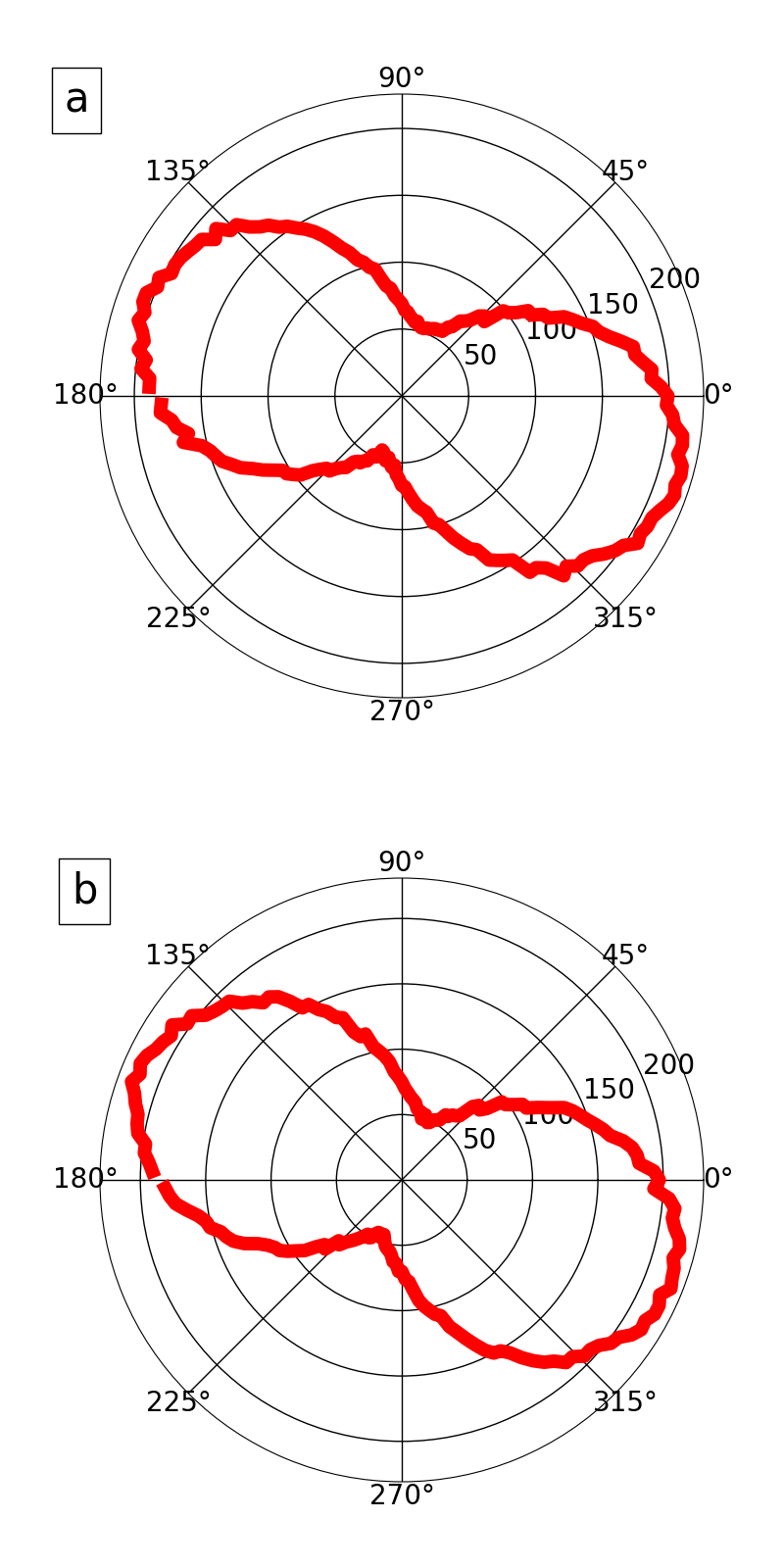}
\caption{\label{fig4} Dependence of the transmitted through the PET film intensity on the rotation angle of the polirizer at $\phi=0$ and normal incidence ($\psi=0$) when using Ti:Sa laser in fs pulse mode (a) and in continuous mode (b)}
\end{figure}

Next, we analyzed the ellipticity of radiation after transmission through the PET film when using femtosecond Ti:Sa master oscillator  both  in continuous (805 nm) and  femtosecond (805 nm with spectral   width 40 nm, pulse duration 40 fs, 0.5 W at 80 MHz repetition rate) modes (see fig \ref{fig4}). In the figure \ref{fig4} the dependence of intensity registered on the angle of polarizer rotation for $\psi=0$ (normal incidence) and $\phi=0$ (at vertical plane) is shown. For the both regimes ellipticity is equal to 0.45. Thus, the PET film could serve for circular polarization obtaining for the femtosecond mode also, and the results obtained for the CW mode correctly describe ellipticity dependence for the femtosecond mode. The only difference is due to the slight difference in wavelengths of diode and Ti:Sa laser radiation.

We also study degradation of the film under action of intense femtosecond radiation. The damage threshold of the PET film was measured. The criterion for the film damage was its darkening due to the two-photon absorption. The film can withstand $10^3$ shots at intensity of $I = 3 \cdot 10^{12}$ W/cm$^2$,  $10^4$ shots -- at  $I = 2.3 \cdot 10^{12}$ W/cm$^2$, and   $>10^6$ shots -- at $I = 1.6 \cdot 10^{11}$ W/cm$^2$, without visible damage. Thus, for a PW laser system and beam diameter of 20 cm the PET film can be used without damaging for quite a long time.

The self action of a femtosecond radiation in the film is also negligible.  Let's evaluate the nonlinear phase shift $B$ mentioned above. The nonlinear refractive index of the PET is\cite{doi:10.1088/1612-2011/12/2/025301} $n_2\approx 5\cdot10^{-20}~\frac{m^2}{W}$  and for 1 TW laser beam 2 cm in diameter at the wavelength of 800 nm the $B$-integral acquired in the film 21.25 $\mu$m thick is 0.026, i.e. much less than unity. Hence  there will be no feeble self-phase modulation or self focusing from such a film.

To demonstrate the possibility of using the PET film to alter the polarization in a high-intensity femtosecond laser-plasma experiment, we measure angular distribution of the second harmonic from the plasma channel sheath in the electron acceleration experiment with a film target.

The  1 TW Ti:Sa laser system at MSU (800 nm, 10 Hz, 50 fs, 50 mJ on target) was used, the maximum vacuum intensity reached  $\approx 5 \cdot 10^{18}$ W/cm$^2$. The target was another PET film  $12~\mu$m thick. Nd:YAG laser (1064 nm, 10 Hz, 200 mJ, 10 ns) with an intensity of up to $5 \cdot 10^{12}$ W/cm$^2$ was utilised to create an expanding plasma plume from the target surface. By changing the delay between the laser pulses in the range from -50 ns to +10 ns with jitter of 1 ns, one can change electron plasma density profile \cite{doi:10.1088/1612-202X/ac6fcb}. To change Ti:Sa laser radiation  polarization the same PET film $21.25~\mu$m thick was used.

To visualize the phenomenon of the second harmonic $2 \omega$ generation from the plasma channel sheath, observed when the plasma channel is formed \cite{doi:10.1103/PhysRevLett.101.045004}, a lens was installed in the direction of the laser pulse propagation. The image was recorded by the CCD camera with an $450\pm20$ interference filter.

\begin{figure}[h!]
\centering
\includegraphics[scale=0.45]{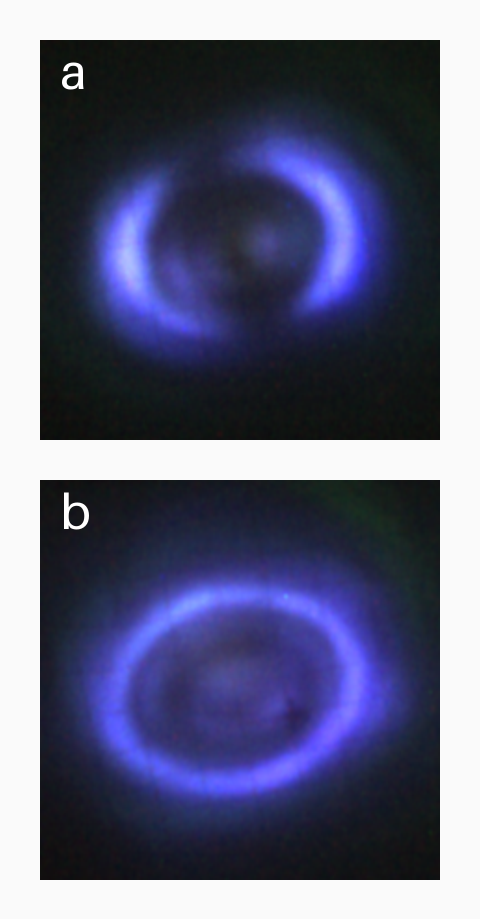}
\caption{\label{fig3} Second harmonic radiation in the experiment. Laser polarisation: linear (a), elliptical with about 0.7 ellipticity (b).}
\end{figure}

The $2 \omega$ radiation in this experimental scheme is emitted as Cherenkov radiation, i.e in an open cone. The laser radiation polarization  determines the angular dependence of the intensity of the second harmonic: the maximum intensity is observed in the plane of laser radiation polarization. Thus, for circularly polarized laser radiation there is a closed ring distribution and for linearly polarized - intermittent ring. This exactly one can see in fig. \ref{fig3}. Thus, the PET film can be used to change the polarization in the experiment with high-intensity laser pulse.

\section{Conclusion}
In this article we have proposed utilisation of the PET film for obtaining a circular polarization of high-intensity wide-band laser radiation. It was demonstrated that a PET film can give laser radiation polarization with ellipticity of up to 0.8. The transmission is $70-80\%$. Within the laser beam size of 2 cm, ellipticity of polarization is stable. PET film can be used for both continuous mode and for femtosecond  mode. The PET film is often used to protect optical equipment. This article demonstrates that its influence on the polarization of laser radiation should not be neglected, and this will affect the interpretation of the experiments results where a PET film was used. PET film can be utilised to analyze the polarization effect in experiments of interaction of high-intensity laser radiation with matter. We demonstrated this using 1 TW Ti:Sa laser radiation. But also the PET film can be used on PW laser system (at a 20 cm diameter, intensity of $ 3 \cdot 10^{12}$ W/cm$^2$ the film is not damaged for $10^3$ shots).

\begin{acknowledgments}
This work was supported by RSCF grant №22-79-10087. E.M. acknowledges Foundation for the advancement of theoretical physics “BASIS” for the financial support.
\end{acknowledgments}

\bibliography{main}

\end{document}